\documentclass[aps,prl,twocolumn,superscriptaddress]{revtex4-2}\usepackage[utf8]{inputenc}

\usepackage{amsmath}

\usepackage{bm}
\usepackage{amssymb,amsfonts,latexsym,fancyhdr,graphicx,epstopdf}
\usepackage[utf8]{inputenc}
\usepackage{soul}
\usepackage{graphicx}

\usepackage[english]{babel}
\usepackage{graphicx}
\usepackage {subfigure}
\usepackage[pdftex,colorlinks=true,allcolors=blue]{hyperref}

\usepackage{xcolor}

\usepackage{verbatim}

\newcommand{\canc}[1]{}

\begin{document}

\title{Enhancing qubit readout with Bayesian Learning}

    \author{F. Cosco}
    \email{francesco.cosco@vtt.fi}
    \affiliation{Quantum algorithms and software, VTT Technical Research Centre of Finland Ltd, Tietotie 3, 02150 Espoo, Finland}
    \author{N. Lo Gullo}
    \email{nicolino.logullo@unical.it}
    \affiliation{Dipartimento di Fisica, Universit\`a della Calabria, 87036 Arcavacata di Rende (CS), Italy}
    \affiliation{INFN, Sezione LNF, gruppo collegato di Cosenza}
    \affiliation{Quantum algorithms and software, VTT Technical Research Centre of Finland Ltd, Tietotie 3, 02150 Espoo, Finland}
    
\begin{abstract}
We introduce an efficient and accurate readout measurement scheme for single and multi-qubit states. Our method uses Bayesian inference to build an assignment probability distribution for each qubit state based on a reference characterization of the detector response functions. This allows us to account for system imperfections and thermal noise within the assignment of the computational basis. 
 We benchmark our protocol on a quantum device with five superconducting qubits, testing initial state preparation for single and two-qubit states and an application of the Bernstein-Vazirani algorithm executed on five qubits. Our method shows a substantial reduction of the readout error and promises advantages for near-term and future quantum devices.
\end{abstract}

\maketitle

\emph{Introduction}--
The promises of quantum computing as a revolutionary technology are being challenged by severe technological limitations in the current hardware. The noise level in the single and two-qubit gates is undoubtedly one such factor \cite{Arute2019}, if not the most important one at the moment.

On the one hand, this prevents the building of fault-tolerant quantum computers and makes the current noisy intermediate-scale quantum (NISQ) processing units of limited use, especially when compared with classical (super-)computers. On the other hand, these limitations are fostering the emergence of new research aiming at exploiting these devices as they are \cite{Preskill2018,Kandala2019,Bharti2022}, not seeing them as a first rough attempt towards a fully fault-tolerant quantum computer. 
Possible solutions come from improving the design of quantum processing units \cite{Marxer2023,li2023quantum} and noise-resilient qubit registers \cite{gonzalez2022,bartling2022,Pedernales2020,you2023scalable} to improving the control electronics, from increasing manufacturing quality \cite{Hyyppa2022} to creating algorithms and software to optimize the data pre/post-processing or recurring to quantum optimal control theory \cite{basilewitsch2019,abdelhafez2020,Bullock2018,werninghaus2021}.

Quantum error mitigation is the general framework grouping all techniques that aim to improve the performances of NISQ processing units \cite{Endo2018,Endo2021,Bultrini2021,Lowe2021,Benjamin2022}, typically by post-processing the data of the quantum computation to decrease noise impact\cite{Strikis2021,Kim2022}, not without some fundamental limitation~\cite{Takagi2022,Kim2023}.

In a quantum processor, errors arise at all different stages of computation alike in a classical one: i) when loading data (state preparation); ii) during data processing (circuit execution); iii) at the moment of the readout (measurement). Most of the focus has been on mitigating errors at the level of the quantum circuit execution, e.g. zero noise extrapolation \cite{Li2017,Temme2017,GiurgicaTiron2020,Bae2021,Krebsbach2022}.
However, several recent works have focused on improving the readout fidelity~\cite{Reed2016,Walter2017,Harvey-Collard2018,Burkard2019,Heinsoo2018,Touzard2019,Urbanek2020,Martinez2020,Lecocq2021,Geller2021,Lin2021,Chen2022,Maurya2022}, thus reducing the error which is still around $\approx 1-2 \%$ in the best cases.
Such a large error alone is enough to prevent from exploiting the potential of quantum computers either in their NISQ version or, even worse, if we aim at error correcting algorithms.

\begin{figure}[!t]
\begin {center}
   \includegraphics[width=\columnwidth]{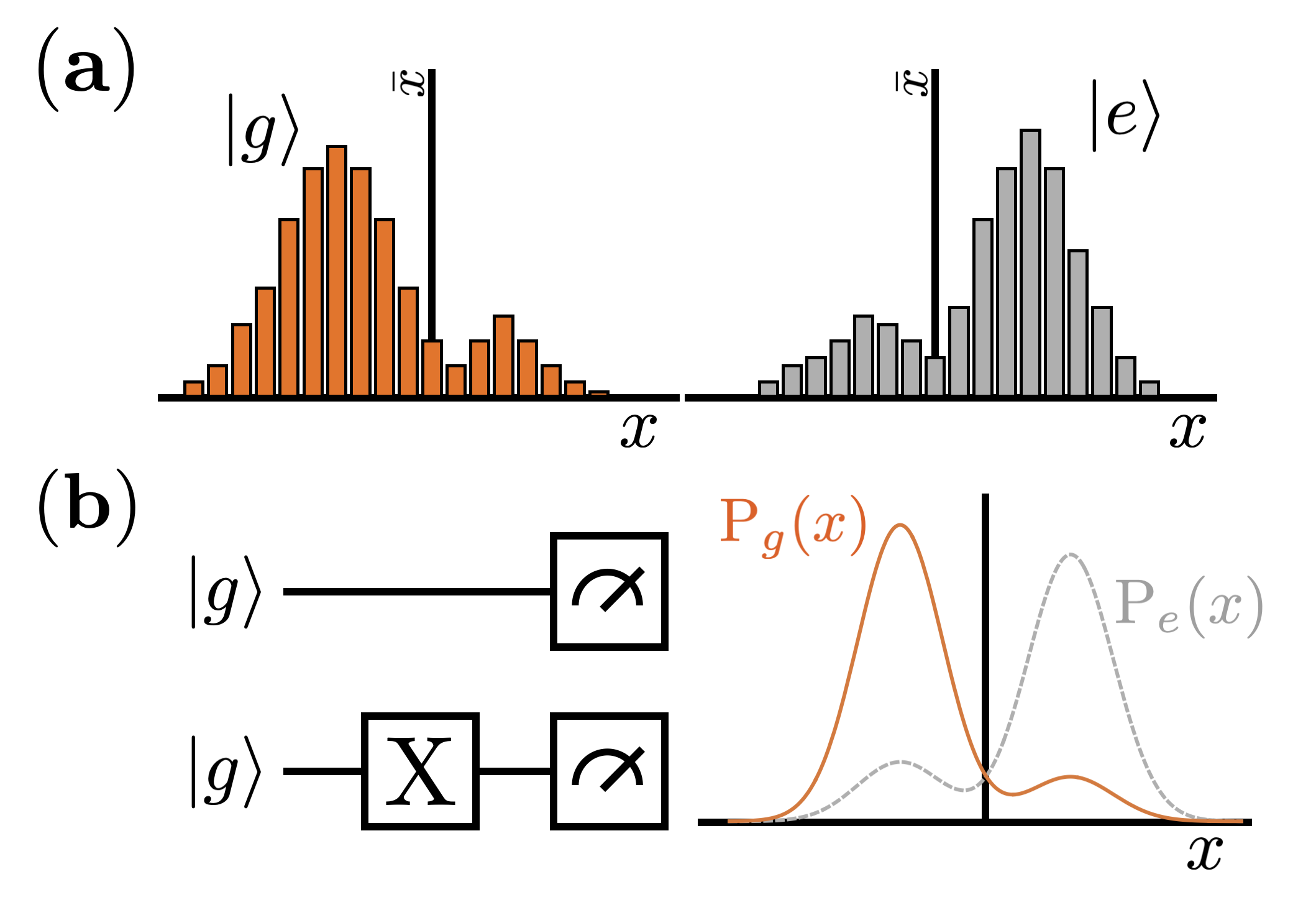}
\end{center}
\caption {(a) Typical qubit-state-dependent response functions. (b) Sketch of the characterisation step: circuits and fit of the response functions $\mathrm P_g(x)$ and $\mathrm P_e(x)$.}
\label {fig:sketch}
\end{figure}


The approaches to improve the readout fidelity include a series of techniques aiming at improving the pulse protocol before the measurement \cite{Ivanov_2023}, finding the optimal readout time \cite{Gambetta2007} for the wanted fidelity of the discrimination or the post-processing of the counts \cite{Maciejewski2020}. 
Methods which manipulate the outcomes statistics, with the inversion of the confusion matrix, may attain the intended results but can yield non-physical values, an issue commonly addressed by incorporating a convex optimizer in the mitigation pipeline \cite{ChenPRA2019}. Additionally, the scalability of matrix inversion can become challenging when dealing with higher dimensions.
Some other proposals to improve the discrimination between the two states, which are gaining attention, are based on machine learning-based techniques \cite{Strikis2021}, either in post-processing \cite{Peters2023} or during the readout protocol \cite{Navarathna2021,Lienhard2022,Azad2022}.

In this work, we introduce a statistical readout framework based on Bayesian inference, inspired by successful application of Bayesian inference in quantum metrology and sensing \cite{Dinani2019,Puebla2021} or parameter estimation for quantum circuits and states~\cite{Teklu2009,Paesani2017,Laverick2021,Smerzi2021,Duffield2022,Leibfried2005,Lundeen2009,Cassemiro_2010}. Our Bayesian Learning Readout (BaLeRO) relies upon two steps: i) characterization of the readout device to build the detector response functions, akin to detector tomography \cite{Maciejewski2020,feito2009}; ii) post-processing data obtained from the circuit execution using a Bayesian update rule and the functions obtained from the first step. Moreover, the post-processing algorithm's input is the raw data from the detector (e.g. quantum analyzer), and the output is some probability distribution for the occupation of the computational basis (e.g. for a single qubit $|0\rangle$ and $|1\rangle$).
Crucially, this approach is not a single-shot discriminator but leverages all measurement data to reconstruct the system density matrix.
One of the advantages of our method is that it avoids the issue of non-physical outcomes and provides a stable foundation for extending the framework to the multi-qubit scenario, avoiding posterior optimization of the mitigated outcomes and noise matrix inversion altogether with an iterative approach. The core concept behind our Bayesian readout is constructing a probability distribution for qubit populations that remains consistent with the measured data, regardless of the system dimension. This is accomplished by leveraging the conditional probabilities that
establish a mathematical and physical connection between each individual measurement and the detector response
functions.
 In what follows, we first introduce the single-qubit framework as a foundational step. Then, we show how to extend the methodology to operate in higher dimensions with a larger number of qubits without suffering the issues arising in the other methods based on the inversion of the noise matrix.

We show that our algorithm can improve the results obtained on actual quantum processing units for several quantum circuits and algorithms. 
Although we primarily focus on superconducting qubit devices, the approach could be applied to other platforms which allow for the reconstruction of the readout device response functions.

\begin{figure}[!t]
\begin {center}
 \includegraphics[width=0.97\columnwidth]{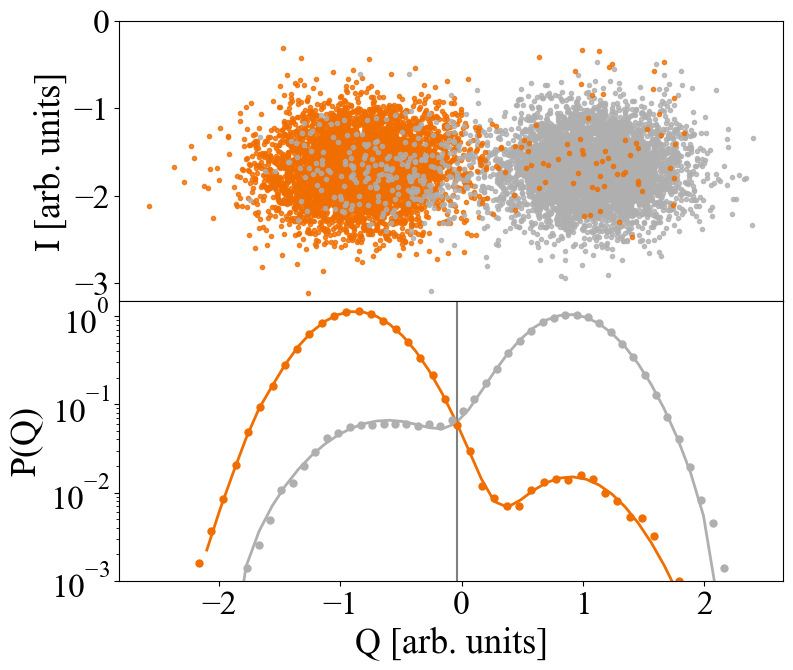}
\end{center}
\caption {Top-panel: Measurement clouds in IQ-plane. Bottom-panel: Projected histograms and fitted response functions of IBM Quito Q4.}

\label {fig:clouds}
\end{figure}

\emph{Embrace the noise}--
A qubit is encoded in a quantum system with at least two distinguishable states, say $| g \rangle$ (ground) and $| e \rangle$ (excited). The readout procedure aims at assigning $0$ or $1$ depending on which of those states the qubit is at the moment of the measurement.
However, detection is not perfect. Instead of getting two distinguished and sharp signals depending on the qubit's two possible states, a detector's typical response function is similar to the one in Fig.~\ref{fig:sketch} {\bf (a)}. The $0$ or $1$ assignment is done through a separatrix which discriminates the state after each measurement. The separatrix is calibrated beforehand and might be either a curve in a two-dimensional space, like the IQ-plane of Fig.~\ref{fig:clouds} (top) for superconducting qubits \cite{Blais2021}, or a threshold in a projected space as in Fig.~\ref{fig:clouds} (bottom).  Hence, the standard readout method assigns either a '0' or '1' based on the relative position of the physical value with respect to the separatrix for each individual measurement. Once all individual measurements are completed, they are aggregated to yield the final count statistics. This approach is prone to a significant error because of the overlap between the two signals, which is caused by all possible noise sources.

To mitigate these issues, we develop a radically different approach to the problem, which \emph{embeds} the base readout noise into the {\it assignment} process.
We achieve the goal in two steps: i) characterization of the detector; ii) post-processing of the data collected after the execution of a generic quantum circuit through a Bayesian-like update rule.
We characterize the detector by performing two simple experiments: reset of the qubit and measurement; flip of the qubit to the
excited state and measurement. The two processes are depicted as quantum circuits in Fig.~\ref{fig:sketch} {\bf (b)} (left).
We repeat these experiments and collect enough data to build two probability distributions $ \mathrm P_g(x)$ and $ \mathrm P_e(x)$. They can be interpreted as the probability distributions of measuring the physical value $x$ (e.g. a voltage) when the qubit is in the ground or excited state.
We construct model fits for these distributions as shown in Fig.~\ref{fig:sketch} {\bf(b)} (right) and in Fig.~\ref{fig:clouds} using bimodal Gaussian distributions. 
We observe a finite interval of values of $x$ for which the two distributions overlap. For these detector outcomes, assessing the qubit's state and assigning the value $0$ or $1$ unambiguously is impossible. This is the primary source of error in the separatrix-based assignment, {\it regardless} of the error's microscopic source.
The second step is now to process the outcomes of any quantum circuit. To formalize our approach, we define two events: (1)-The detector returns $x$ when measuring the state of the qubit
(2)- the qubit's density matrix populations are $\{\rho_g\, \rho_e\}$, i.e. the probability that a measurement performed in the computational basis will find the system in $|0\rangle$ or $|1\rangle$ respectively. These two events are dependent, and we leverage their statistical correlation to process the readout outcomes. Using the model distributions, we define the conditional probability for the detector measuring $x$ given that the qubit's state is characterised  by the pair $\{\rho_g,\rho_e\}$:
\begin{equation}
\mathrm{P} (x|\{\rho_g, \rho_e\}) = \frac{\rho_g \mathrm P_g(x) +\rho_e \mathrm P_e(x)}{\rho_g+\rho_e}.
\label{cond-prob}
\end{equation}

\begin{figure*}[!t]
\begin {center}
\includegraphics[width=1.95\columnwidth]{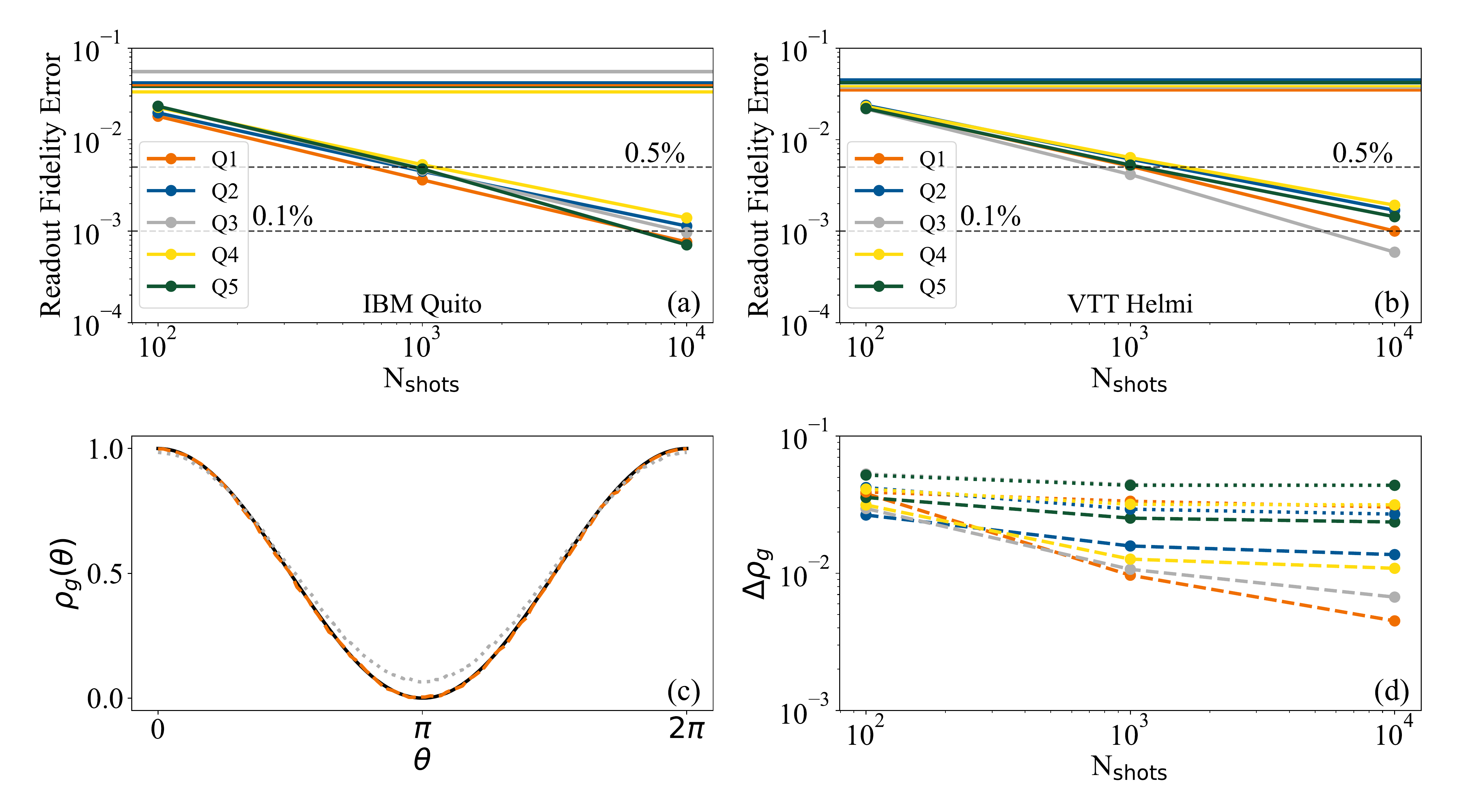}
   
\end{center}
\caption {(a)-(b): Readout fidelity error using BaLeRO as a function of the number of shots for each qubit on IBM Quito and VTT Helmi. The horizontal lines correspond to the errors reported by the service provider \cite{Supp} (in the appropriate color scheme for each qubit).
(c): Ground-state population of Q1 (IBM Quito) after applying a $R_y(\theta)$ gate. The solid black line corresponds to the ideal population, the dotted gray one to the counts estimate and the dashed orange to the BaLeRO estimate with $10^4$ shots. (d): Average ground-state population error, $\Delta \rho_{g}=1/(2\pi) \int \, d\theta {|\cos (\theta/2)^2 -\rho_g (\theta)|}$ as a function of the number of shots for each qubit (IBM Quito). Dotted and dashed lines refer to the counts and BaLeRO estimates respectively (colour scheme as in (a)).}
\label {fig:fass}
\end{figure*}

Likewise, from Bayes' theorem, we can formally write the conditional probability for the qubit's state populations given a measured value of $x$:
\begin{equation}
\mathrm P(\{\rho_g,\rho_e\}|x)=\frac{\mathrm P(x|\{\rho_g,\rho_e\})\mathrm P(\{\rho_g,\rho_e\})}{\mathrm P(x)}.
\end{equation}
Contrarily to the common approach, our algorithm leverages Eq. \eqref{cond-prob} and the full set of detector's measurement outcomes $\mathbf x = x^{(1)},...,x^{(\mathrm N_{\mathrm{shots}})}$ 
trough a Bayesian update rule:
\begin{eqnarray}
&&\mathrm P(\{\rho_g,\rho_e\}|x^{(n)})\propto \mathrm P(x^{(n)}|\{\rho_g,\rho_e\})\mathrm P(\{\rho_g,\rho_e\})\nonumber\\
&=& \left [ \rho_g \mathrm P_g(x^{(n)}) +\rho_e \mathrm P_e(x^{(n)}) \right ]\mathrm P(\{\rho_g,\rho_e\})\\
&\approx& \left [ \rho_g \mathrm P_g(x^{(n)}) +\rho_e \mathrm P_e(x^{(n)}) \right ]\mathrm P(\{\rho_g, \rho_e\}| x^{(n-1)})\nonumber
\label{bayes-update}
\end{eqnarray}
where $\mathrm P(\{\rho_g, \rho_e\})$ is the prior probability distribution and it is updated after each iteration step with the posterior probability distribution $\mathrm P(\{\rho_g, \rho_e\}| x^{(n-1)})$ at the previous step. When assuming no prior knowledge about the system state, we start the iteration  with a uniform probability distribution $\mathrm P(\{\rho_g, \rho_e\}| x^{(0)})$.

Within this framework, we work with a probability distribution defined on the plane $\{\rho_g,\rho_e\}$, in the region which satisfies $\rho_g+\rho_e =1$. The result is the probability distribution $\mathrm P(\{\rho_g, \rho_e\}| x^{(\mathrm N_{\mathrm {shots}})})$ which represents the probability distribution for the qubit's state populations coherent with the collected raw measurement data. With this we can compute the average value of {\it any} single qubit operator projected onto the computational basis or simply recover an estimate for the qubit populations via integration, i.e. 
$\rho^{\mathrm{est}}_{g/e} = \int \mathrm d \rho_g d \rho_e  \mathrm \rho_{g/e} \mathrm P(\{\rho_g, \rho_e\}| x^{(\mathrm N_{\mathrm {shots}})})$.

The method is not limited to single-qubit readout, but it can be applied to any multi-qubit system following similar steps. Given a system of $\mathrm {N_q}$ qubits, and $\mathrm {N_q}$ independent detectors registering the set of measurement outcomes $\mathbf x =x_1, ...x_{\mathrm{N_q}}$ for each shot, the multi-dimensional response function depends now on the population of each multi-qubit state and is written as
\begin{equation}
\mathrm{P} (\mathbf x|\bm \rho \equiv \rho_0,...,\rho_{2^{\mathrm{N_q}}-1}) =\sum_i \rho_i \mathrm P_i(\mathbf x),
\end{equation}
where $\rho_i$ is the population of the $i-$th state, e.g. $| \rho_0 \rangle = | 0...0 \rangle$, $| \rho_1 \rangle = | 0...1 \rangle$ up to $| \rho_{2^{\mathrm{N_q}}-1} \rangle = | 1...1 \rangle$, and $\mathrm P_0(\mathbf x)=\mathrm P_g(x_1) \mathrm P_g(x_2)... \mathrm P_g(x_{\mathrm{N_q}})$, $\mathrm P_1(\mathbf x)=\mathrm P_g(x_1) \mathrm P_g(x_2)... \mathrm P_e(x_{\mathrm{N_q}})$, and so forth.
The posterior probability distribution for the population of each state can be then reconstructed in a sequence of single-shot measurements following the Bayes’ theorem as previously introduced
\begin{equation}
\mathrm P(\bm \rho | \mathbf x^{(n)}) \propto \sum_i \rho_i \mathrm P_i(\mathbf x^{(n)}) \mathrm  P(\bm \rho| \mathbf x^{(n-1)}),
\label{eq:many-qubit-bayes}
\end{equation}
 within the parameters space which satisfies $\sum_i \rho_i = 1$.
It is worth mentioning that our approach shares some similarities with the Iterative Bayesian Unfolding, which is
a form of regularized matrix inversion applied to the response matrix \cite{Nachman2020,Hicks2021}. Here, however, we bypass the faulty state-assignment step and work on the continuous space of the physical measurement.

\emph{Improvement of readout fidelities}--
We test our readout scheme on two five-qubits quantum computers, the IBM Quito and the VTT Helmi (QPU from IQM).
We start by reconstructing the reference response functions by running the circuits depicted in  Fig.  \ref{fig:sketch} (b) for $\mathrm N_{\mathrm {shots}}=10^5$. 
A typical outcome is the one displayed in Fig.~\ref{fig:clouds}(top-panel) where each measurement is represented by a dot in the I-Q plane. The clouds are then often rotated and projected onto one of the axis to obtain the distributions shown in Fig.~\ref{fig:clouds} (bottom-panel). We use these histograms to fit the probability distributions $\mathrm P_g(x)$ and $\mathrm P_e(x)$ using bimodal Gaussians.
We repeat the characterisation procedure for each qubit on the device and then use the obtained response functions to analyse the assignment fidelity in subsequent experiments. In Fig. \ref{fig:fass} (a)-(b), we show the average error related to the readout fidelity for each qubit as a function of the number of shots using BaLeRO. Interestingly, we are able to attain a readout error below $0.5 \%$ already below $10^4$ shots, significantly improving the readout error reported by the service provider which is typically of the order of $\lesssim 5\%$. Strikingly, with our protocol, the readout fidelity improves with an increasing number of shots. While with usual readout methods, the readout fidelity is mainly insensitive to the number of repetitions and depends solely on the device's calibration. It is worth noting that a more in-depth analysis of the QPU setup has the potential to result in more advanced models for the detector response functions, ultimately leading to higher levels of fidelity.
\begin{figure*}[!t]
\begin {center}
  \includegraphics[width=2.2\columnwidth]{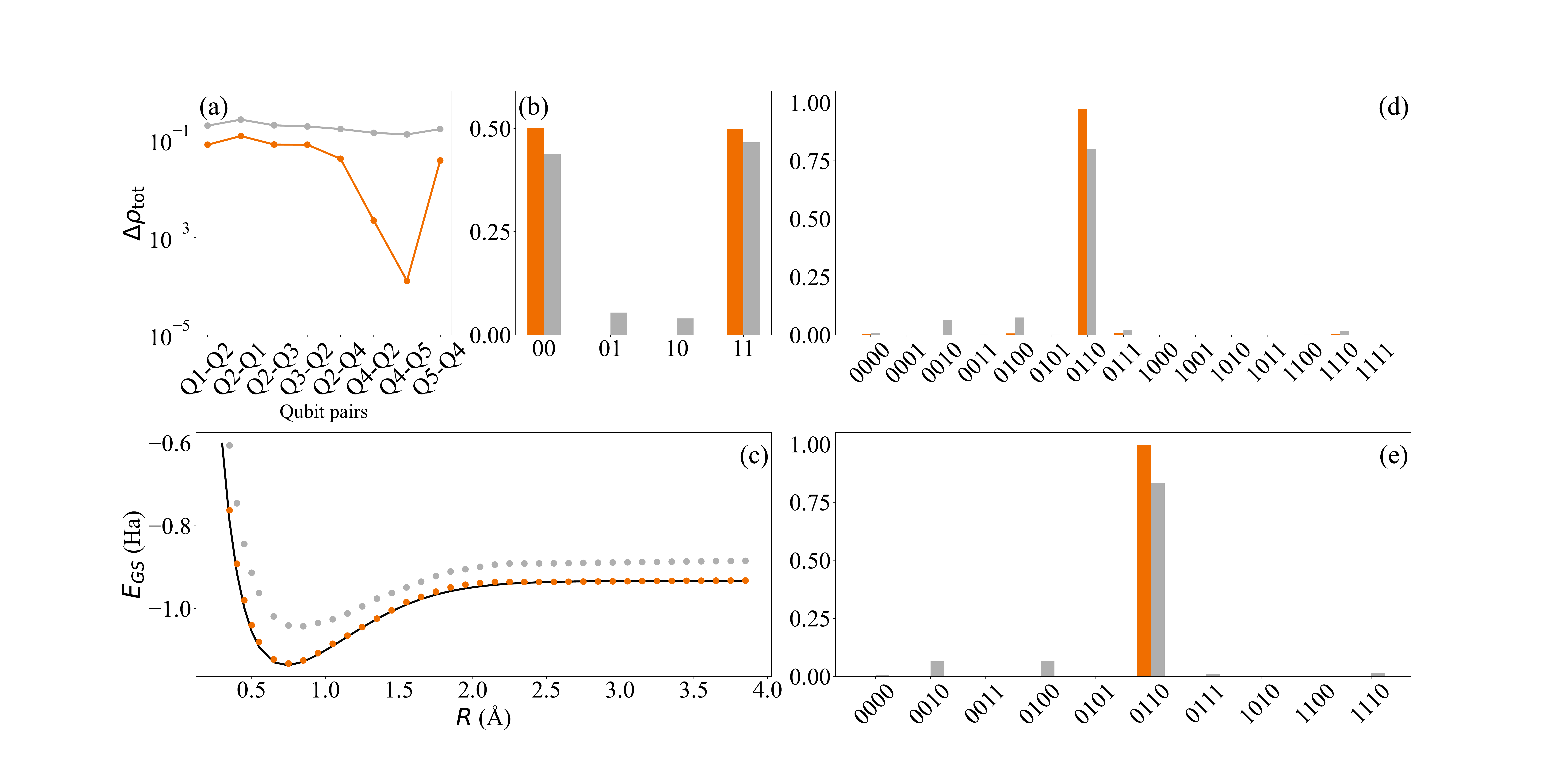}
\end{center}
\caption {(a): Total population error $\Delta \rho_{\mathrm{tot}}=\sum_i {|\rho^{\mathrm{exact}}_i -\rho^{\mathrm{estimated}}_i|}$  for each qubit pair (the first qubit is the control and the second one is the target in the CNOT), and populations of the pair Q4-Q2 (b), for the Bell state  $|\Phi^+\rangle$ measured using BaLeRO (orange) and count statistics (gray), with $10^4$ shots. (c): $H_2$ ground-state energy  measured using BaLeRO (orange) and count statistics (gray) using $10^3$ shots on pair Q1-Q2, compared with the theoretical prediction (black). Populations measured using BaLeRO (orange) and count statistics (gray) in the four qubits system prepared in the state "0110" as output of the Bernstein-Vazirani Algorithm in (d) and state preparation trough qubit flips in (e), both measured with $10^4$ shots. All the above demonstrations were performed on IBM Quito.}
\label {fig:bell}
\end{figure*}

To show the benefits of our approach to the execution of a quantum circuit, we start by considering the application of a single-qubit gate. We choose an arbitrary rotation around the Y axis, i. e. a $R_y(\theta)$ gate, which results in the final populations $\rho_g=\cos (\theta /2)^2$ and $\rho_e=\sin (\theta /2)^2$ when applied to the ground state.  Interestingly, our method outperforms the standard readout scheme as it follows more closely the theoretical prediction, as shown for one of the qubits in Fig.~\ref{fig:fass} (c). In Fig.~\ref{fig:fass} (d), we quantify these improvements by displaying the average error on the rotated ground state population as a function of the number of shots.  Our protocol achieves a smaller error than the standard (non-mitigated) case for any number of shots and each qubit. 
Although the error for the non-mitigated case shows some improvement with an increasing number of shots, saturating around $10^4$ shots, our approach always offers a dramatic improvement for all qubits.

\emph{Multi-qubit use-cases and scalability}--
A typical quantum algorithm requires to perform two-qubit operations and compute mean values of multi-qubit observables. In this section, we show how to employ BaLeRO in such cases and apply it to specific algorithms.
Therefore, as the first and simplest extension, let us consider the preparation of the Bell state $|\Phi^+\rangle=1/\sqrt(2)(|00\rangle+|11\rangle)$, which requires the application of a  Hadamard gate and a CNOT. Thus, we prepare such a state on each directly connected qubit pair, and, as a figure of merit, we calculate the total population error, i.e. deviation from the ideal case, and compare the one resulting from our readout method with the one from the standard scheme in  Fig.~\ref{fig:bell} (a). For each qubit pair, our protocol outperforms the count statistics. As a qualitative example, in Fig.~\ref{fig:bell} (b), we  compare the populations of one of the experiments and see how our protocol greatly reduces the populations from the erroneous states, getting closer to the ideal two-qubit state.

In many practical applications, having access to an accurate description of the two-qubit density matrix is crucial. For example, calculating ground state properties of many-body systems, such as spin -chains, requires, at most, two-body operators. This also applies to simple molecular systems or ferromagnetic models \cite{Tazhigulov2022}, which after "qubitization" have a Hamiltonian with at most two-body operators. As proof of principle, we consider the $H_2$ molecule, whose ground state energy can be calculated through a simplified two-qubit system  \cite{Omalley2016,Ganzhorn2019}. In Fig. \ref {fig:bell} (c), we show the lower energy extracted  using the state preparation ansatz, and Hamiltonian data, of \cite{Urbanek2020} for $\mathrm N_{\mathrm {shots}}=10^3$. Again, we see how our method outperforms the non-mitigated one confirming that the error on the readout is a significant limiting factor in current devices.

Up to two-qubit quantities, the post-processing algorithm is computationally inexpensive. However, the computational cost increases rapidly, actually exponentially, with the number of qubits making it unfeasible. For the sake of clarity, for a system with ${\mathrm N_{\mathrm q}}$ qubits, Eq.~\eqref {eq:many-qubit-bayes} requires handling, at each iteration step of the learning, a multi-parameter probability distribution $\mathrm P( \bm \rho|\mathbf{x}^{(n)})$ where each variable, i.e. the  $2^{{\mathrm N_{\mathrm q}}}$ different populations $\rho_i$, is sampled in the interval $[0,1]$. If each interval is sampled simultaneously with $n_p$ points, the total probability distribution would require a mesh-grid with $n_p^{2^{{\mathrm N_{\mathrm q}}}}$ points, albeit the normalisation condition reduces the effective number of independent coordinates.

Therefore, it appears quite clear that in the multi-qubit scenario our approach is resource intensive and cannot be used as it is, and it requires to be adapted. Thus, we follow a heuristic approach, and instead of working with the complete multi-parameter probability distribution, we demote all the populations except two, namely $\rho_i$ and $\rho_j$, to constant estimates $R_k$, and define the conditional probability function
\begin{equation}
\mathrm{P} (\mathbf x|\mathbf \rho_i,\rho_j) =\rho_i \mathrm P_i(\mathbf x)+\rho_j \mathrm P_j(\mathbf x)+\sum_{k \neq i,j}R_k \mathrm P_k(\mathbf x).
\end{equation}
Using this, we apply the Bayesian cycle to build a posterior distribution for the population pair in exam, which is then used to compute some new estimates $R_i$ and $R_j$. The algorithm needs to be repeated for each population pair $(\rho_i, \rho_j)$ and iterated to reach convergence.
The convergence criterion can be chosen at will; a simple rule of thumb is to stop the iterations if the change in the posterior probabilities or the obtained populations does not change with respect to the previous iteration by an amount $\epsilon$.

To further lighten the computational cost, we  reduce the number of parameters $\rho_i$ included in the optimization procedure. The strategy is to use the initial estimates $R_k$, which at the beginning can be
the result obtained via the count statistics, to drop states which have a negligible population according to a pre-defined threshold. In our test we drop from the optimization the states for which we do not get initial counts.

While a highly entangled many-qubit state might still be difficult to reconstruct, we apply our protocol to a quantum algorithm whose theoretical output is a single bitstring, i.e. the Bernstein-Vazirani algorithm \cite{Bernstein1997,Simon1997,Lidar_2023}. In Fig. \ref{fig:bell} (d), we compare the populations measured using our approach, with parameter region restriction and Bayesian update applied to population pairs, and the count statistics. We apply the Bernstein-Vazirani algorithm on a five qubits system to obtain the four qubits string "0110" as an output. Our protocol increases the population of the correct output string, reducing the error from $\sim 20\%$ to $\sim 3\%$. The remaining erroneous populations are mainly due to imperfections in the qubit gates.
To confirm it, we prepare the same string, "0110", with a simpler circuit, i.e. by flipping the qubits. The count statistics shows errors of the same order of magnitude as in the execution of the Bernstein-Vazirani algorithm (Fig.~\ref{fig:bell} (e)), while with BaLeRO we get very close to the expected outcome, reducing the error from $\sim 16\%$ to $\sim 0.15\%$. {It is worth mentioning that when measuring multiple qubits simultaneously, the response functions can be affected by a form of correlated noise, i.e. the readout signal of one qubit might be affected by the simultaneous measurement of another. In principle, this correlated noise might be accounted for when calibrating the detector response functions at the cost of a more demanding calibration step. However, in our multi-qubit applications, where we performed two-qubit and four-qubit measurements, we have seen a considerable improvement in the readout fidelity, hinting that if the crosstalk is not large, our scheme can correct it even when the fitting functions are built via uncorrelated single qubit measurements only.}

\emph{Conclusions}--
We have introduced an efficient readout scheme, BaLeRO, which significantly improves the accuracy of the readout step in quantum computers. The algorithm employs Bayesian inference to build a probability distribution for each qubit-state population based on a reference characterisation of the detector response functions, thus including noise and imperfections in the post-processing of the raw measurement outcomes.  Importantly, by design, our Bayesian framework avoids assigning unphysical values and guarantees that the mitigated outcomes are consistent with the experimental data.
We tested BaLeRO on two quantum computers with five superconducting qubits, and we have demonstrated an accuracy improvement in terms of single- and up to four-qubit readout. 
Although we applied the protocol to a superconducting qubit architecture, BaLeRO applies to other platforms for digital quantum computation. 
It is worth stressing that BaLeRO is not a single-shot post-processing protocol based on a discriminator to assign "0" or "1" to each measurement. Standard approaches rely on the quality of the readout protocol itself while improving the choice of the discriminator, making the final results are strongly related to the quality of the calibration of the machine at the time of computation.
BaLeRO instead incorporates the readout noise in the post-processing, improving the statistical reconstruction of the quantum state coherent with the actual measurement. Therefore when it comes to the computation of mean values of Pauli strings, Bayesian inference offers a robust and more reliable alternative.
Finally, while already proving a reduction of the readout error as it is, embedding our scheme within the calibration process of each gate might further boost the performance of NISQ and future devices. 

\begin{acknowledgments}
We acknowledge the use of IBM Quantum services for this work. The views expressed are those of the authors, and do not reflect the official policy or position of IBM or the IBM Quantum team.
\end{acknowledgments}

\bibliographystyle{apsrev4-2.bst}
\bibliography{biblio}


\end{document}